\documentclass[runningheads]{llncs}
\usepackage{amssymb}
\usepackage{graphicx}

\usepackage{color}
\usepackage{soul}
\usepackage{tikz}
\usepackage{amsmath}
\usepackage{amssymb}
\usepackage{mathtools}
\usepackage{pgfplots}
\usetikzlibrary{patterns}
\usetikzlibrary{calc}
\usetikzlibrary{pgfplots.groupplots}
\usetikzlibrary{backgrounds}
\usepackage[vlined,linesnumbered]{algorithm2e}
\definecolor{brown}{rgb}{.5,.15,.15}
\usepackage{setspace}

\begin{document}
\title{A Timer-Augmented Cost Function for Load Balanced DSMC}
%
%
\author{William McDoniel\orcidID{0000-0002-5890-8543} \and
Paolo Bientinesi}
\authorrunning{W. McDoniel and P. Bientinesi}
%
\institute{
	RWTH Aachen University,
	Aachen, Germany 52062\\
	E-Mail: \texttt{mcdoniel@aices.rwth-aachen.de}
}

\maketitle              

\begin{abstract}
Due to a hard dependency between time steps, large-scale simulations of gas using the Direct Simulation Monte Carlo
(DSMC) method proceed at the pace of the slowest processor. Scalability is therefore achievable only by ensuring that the work done each time step is as evenly apportioned among the processors as possible.  Furthermore, as the
  simulated system evolves, the load shifts, and thus this load-balancing typically needs to be performed multiple times over the course of a simulation.
Common methods generally use either crude performance models or processor-level timers.  We combine both to create a timer-augmented cost function which both converges quickly and yields well-balanced processor decompositions.  When compared to a particle-based performance model alone, our method achieves 2x speedup at steady-state on up to 1024 processors for a test case consisting of a Mach 9 argon jet impacting a solid wall.

\end{abstract}
\section{Introduction}

For the simulation of rarefied gas flows, where collisions between
molecules are both important and sufficiently rare that the gas cannot be treated
as a continuum, the
approach of choice is Direct Simulation Monte Carlo (DSMC) \cite{bird2013dsmc}.
DSMC finds applications across many fields, spanning a huge range of time and
length scales, and treating a wide variety of physics, from flow in and around
microelectromechanical systems (MEMS)~\cite{karniadakis2006microflows}, to the highly reactive flows around a
spacecraft during atmospheric re-entry~\cite{moss1985direct}, to entire planetary atmospheres \cite{MCDONIEL_ATMO}, and even extending to the evolution of solar systems and galaxies~\cite{weinberg2014direct}.
DSMC is a particle-based method where the interactions among individual computational molecules are simulated over time.  A major challenge in running 3D,
time-varying DSMC simulations on supercomputers is load balancing, which is
not a simple matter of assigning or scheduling independent tasks until all are
finished.  In this paper, we propose and test a new method for estimating the
computational load in regions of an ongoing DSMC simulation, which is
  used to divide the physical domain among processors such that each has to do a roughly equal amount of computation.

When run on many cores, each process in a DSMC simulation owns a region of space and all of the particles in it.  The simulation proceeds in time steps, repeatedly executing three actions.  First, all particles move a distance proportional to their velocities.  Second, particles which have moved to regions owned by different processors are communicated.  Third, pairs of nearby particles are tested for collisions and may (depending on random number draws) collide.  In order to determine the effect of collisions on the gas, processors must know about all of the local particles.  Effectively, this introduces a dependency between the movement of particles on all processors and the collisions on a single processor.  All of the processors must proceed through the simulation synchronously, and so performance degrades if one processor is doing more computation per time step than others, since all of the others will be forced to wait for it each time step.  A single processor that takes twice as long as the others to simulate a time step will cause the whole simulation to proceed half as quickly as it otherwise might.

The load balancing of a simulation in which processes own regions of space in a larger domain boils down to the determination of where processor boundaries should be drawn, with the goal of dividing the domain into regions that require roughly equal amounts of
computation each time step.
Almost all methods for load balancing physics simulations rely on a ``cost
function'', which may be implicit in a more complex balancing scheme, or explicitly computed and used as the basis for decomposing the domain.  A
cost function maps points in space to estimated computational load; a simulation is load-balanced if the integral of the (accurate) cost function over a processor's subdomain is the same for all processors.  Cost functions can be produced from performance models, by analyzing the method.  Sometimes this is straightforward.  For example,
many finite difference or finite element solvers predictably do work
proportional to the number of elements or grid points; consequently, such
simulations can be load balanced by evenly apportioning elements to
processors. By contrast, the computational cost of DSMC is hard to
model. Because of this, load balancers tend to either use approximate models
(which have errors and so lead to imbalance) or ignore the details of the
method entirely and look only at actual time spent by different processors
(thus producing a coarse cost function).
We propose to combine both sorts of estimates to achieve better balance than either can by itself.

\section{DSMC}

DSMC is a stochastic, particle-based method which has been in use for decades \cite{bird1976molecular}. The main idea is to use a relatively small number of computational particles (millions to billions) to represent the very large number of real particles in a macroscopic system, where real particle densities can easily be $10^{20}$ particles per cubic meter or larger, and the system could be an entire planet's atmosphere or more.  ``Particles'' are generally molecules (as with the test case in this paper) but may be dust grains, ions, or even electrons.  In DSMC, each computational particle behaves much like a real particle most of the time, e.g., moving in response to external forces.  Collisions between computational particles have a random element.  Random number draws determine post-collision velocities and energies, whether or not chemical reactions occur when particles collide, and even whether or not a collision between two particles occurs at all.  Because each computational particle's properties change over time in the way that a random real particle's might, a relatively tiny number of computational particles can capture the statistical properties of the real flow.

DSMC produces approximate solutions to the Boltzmann equation.  Unlike the dense (often liquid) flows for which molecular dynamics is suited, in a rarefied gas molecules are almost always so far from their nearest neighbors that intermolecular forces are essentially nonexistent.  Molecules interact with other molecules for relatively short amounts of time in between long periods of ballistic motion.  Using the dilute gas approximation, DSMC treats these interactions as instantaneous pair-wise collisions which can be de-coupled from molecular motion within a time step.  As a particle-based method, rather than a differential equation
solver, DSMC is also highly extensible, and modular physics packages can be
quickly implemented, making it easy to use for many different kinds of
problems. DSMC is expensive relative to traditional partial differential equation solvers for
fluid flow, and becomes more so as flow densities increase. Still, the method
is used because it is much more accurate than traditional solvers at low
densities (more precisely, it is more accurate when the mean free path between collisions is not small relative to other length scales in the problem).

A major challenge to load balancing in DSMC is that, unlike with many
continuum finite element methods, it is difficult to model its cost (the difficulty of load-balancing DSMC is also discussed in \cite{IVANOV1997485}).  The
amount of computation performed scales roughly linearly with the number of
particles, but this is not an exact relation.  Most particles need to be moved once per time step, and this is often simple. Most particles are moved by multiplying their velocities by the time step size and adding the result to their positions. But particles which are near a surface or domain boundary may impact it and bounce off, in which case the code must find when and where they intersected the surface and then compute a new move from that point with the remaining time.

Not only are some particles easier to move than others, the number of collisions that a processor needs to compute depends on many factors.  DSMC domains are typically divided into a grid, and the grid cells are used to locate neighbors which are potential collision partners.  The number of collisions to perform in a cell is (on average) proportional to the square of the number of particles in the cell and to the time step size.  When collisions are a significant factor in a simulation's cost, high-density regions are often more expensive than low-density regions, even when both have the same total number of particles.  The number and type of collisions actually performed depends further on local temperature, on the species present, on which chemical reactions are being considered, etc.

Furthermore, particle creation (for inflow at a boundary, for example) might also be a significant cost.  Created particles need to be assigned some initial properties, like velocity, which are typically sampled from probability distributions.

The computational cost of each of these major functions of a DSMC code is difficult to model by itself, and the relative importance of each depends on specific features of the
problem being simulated. In general, it is not feasible to develop a ``one size fits all model'' that predicts computational cost from the state of a simulation.

\section{Load Balancing}

We conceive of load balancing as a two step process.  The first step is to
find a way to predict whether a proposed domain decomposition will prove to be
well-balanced, or at least better balanced than the current decomposition (for
dynamic balancing).  This is the purpose of a cost function, which can be applied to the space owned by a processor to obtain an estimate of the computational load associated with that space; a simulation is well-balanced if all processors are performing roughly equal amounts of work.
The focus of this paper is on this first step: We want to provide a better estimate of the simulation's true cost function.
The second step is
to actually assign parts of the simulation domain to each processor such that
the new decomposition is predicted to be balanced by this cost function.   For this step, there exist many
techniques for splitting up a domain which are suitable for DSMC.  This is a space partitioning problem, and we stress that the two steps are in general separable.  Many different kinds of cost function can be used as input for a given decomposition algorithm (as in this paper), and a given type of cost function can be used with many different decomposition algorithms.  For this work, we implemented a recursive coordinate bisection (RCB) algorithm \cite{berger1987partitioning} to test various cost functions:  Given a map of computational load in a 3D domain, we cut in the longest dimension so that half of the work is in each new subdomain;  this is applied recursively until one subdomain can be assigned to each processor.

While we will not discuss the partitioning problem in detail, it is important
to recognize that it can be expensive.  This is unimportant for a static
  problem, since the balancing must only be performed once, but time-varying simulations will need to be periodically
re-balanced.  Many methods of re-partitioning a domain will often produce a new processor map that has little overlap
with the old one.  That is, after a repartitioning many processors will own
entirely different regions of space. When this happens, there is a large
amount of communication as many (and often most) of the particles in the
domain will need to be sent to new processors, and computation must wait until a processor is given the particles contained within the region of space it now owns. 
Therefore it is undesirable to load balance too frequently, and it is important to have an accurate cost function which does not require multiple balance iterations to obtain a good result.

\subsection{State of the Art}\label{sec:state-of-the-art}

We now briefly discuss the four methods\footnote{the names of the methods are our own labels for them} in common use for load balancing parallel DSMC simulations, highlighting the advantages and disadvantages of each.

\begin{enumerate}
\item Random scattering (e.g., \cite{IVANOV1997485}): Since DSMC performance is hard to
  model and load-balancing itself can be costly, this method seeks to quickly
  and cheaply balance a simulation by dividing the domain up into many small
  elements and then simply randomly assigning elements to processors, ignoring
  topology, adjacency, etc.  Each processor will own a large number of often
  non-contiguous elements.  If the elements are small relative to length
  scales in the problem, it is likely that each processor is doing a roughly
  equal amount of work.  This method might not even require re-balancing for
  dynamic problems, though re-balancing can be necessary if the collision grid changes.
  However, these benefits come with a significant drawback.  By assigning small, non-contiguous chunks of space to each processor, random scattering drastically increases the number of particles that move between processors each time step.  Each processor also neighbors a very large number of other processors (likely neighboring almost \textit{every} other processor), and it is difficult to determine which processor owns a particular point in space.
  
\item Cell Timers (e.g., \cite{nance1994parallel}): One solution to the problem of not knowing
  where the load is is just to measure it.  By inserting timers into a code at
  the cell level, one obtains a resolved map of load in the simulation.
  However, this is difficult to do because many computationally expensive
  parts of a DSMC simulation are not naturally performed with an awareness of
  the grid structure.  The grid is irrelevant to moving particles around, for example,
  and typically only matters for finding collision partners.  Additional indexing to keep track of which cells particles are in may be required, with a very large number of cell-level timers
  turning on and off many times each time step.
	
	\item Processor Timers: A popular class of load-balancer abstracts away almost all of the details of the method and just looks to see which processors are taking longer than others.  Some distribution of the load inside each processor is assumed (typically uniform) and boundaries are periodically redrawn to attempt to achieve a balanced simulation.  Because the map of load as a function of space is very coarse, this method might require multiple iterations to achieve balance, even given a static simulation.  It can also become unstable, and so is typically integrated into a partitioning scheme which will move small amounts of space between processors very frequently (as in \cite{IVANOV1997485}, \cite{OLSON20102063}, or \cite{taylor1996concurrent}).
	
	\item Particle Balancing (perhaps first applied to DSMC in \cite{nance1994parallel} and used in codes like UT's PLANET \cite{MCDONIEL2015251} or Sandia's SPARTA [http://sparta.sandia.gov]): Another popular balancing method uses particle count as a proxy for load.  Processor boundaries are drawn such that each processor owns space containing roughly equal numbers of particles.  This is essentially just a crude performance model, and it often works well for small numbers of processors.  Particle count is also an excellent proxy for memory requirements, and so this method naturally helps ensure that memory use is balanced as well.  However, because this method only approximates the true load, error can yield significant imbalance, especially for simulations using many processors.
\end{enumerate}

The last three of these methods all implicitly or explicitly depend on a cost function.  They intend to balance the simulation by assigning space to processors such that the integral of the cost function is the same over each processor's subdomain.  Two try to measure this cost function, finely or coarsely, and one estimates it with a performance model.

The processor timer and particle balancing methods are convenient and easy to
implement, but error in their estimated cost functions can be a significant problem when using a large number of processors.  This is easily seen with the particle balancing method.  Suppose that there is a small region of the domain where the load is twice as large as would be expected from the number of particles present in it.  Perhaps there is a lot of particle creation occurring here, or there is some complex geometry in the flow and moving the particles takes longer here as they interact with the object, or this is a very hot region and expensive chemical reactions are more likely to occur.  With a small number of processors, particle balancing will still yield a satisfactory result.  The load-intense region is small, and so the extra work done by the processor which owns it will be negligible compared to the work it does in surrounding regions where the estimated cost function works well.  The ratio of the work this processor does to the work some other processor does will still be near unity.  However, as the number of processors increases, eventually there will be a processor which \textit{only} owns part of this load-intense region.  This processor is assigned twice as much work per time step as the typical processor, and the simulation will proceed at roughly 50\% efficiency.

Meanwhile, the processor timer method is blind to the distribution of load within
each processor -- it assumes a uniform cost function inside each processor's subdomain.  This causes problems at high processor
counts because it will often have to perform multiple balancing passes to
achieve a good result.  It is even possible for it to produce a less-balanced
partitioning than the one it started from.  This is an
overshooting problem, and can be addressed with damping terms or other schemes for shifting processor boundaries slowly over multiple passes, but
these worsen its performance in the typical case and lead to the method
requiring multiple iterations to balance even a flow with very little spatial variation.

We propose to combine the best features of these two methods to produce a hybrid timer-augmented cost function, mitigating the individual drawbacks of the particle balancing and processor timer methods while still being cheap to compute and simple to implement.  We will demonstrate a clear improvement over particle balancing at steady state and show that the hybrid method out-performs a simple implementation of the timer-based method for our test case.

\section{Timer-Augmented Cost Function}
The chief advantage of particle balancing is its quick convergence to a good-enough solution, while a
processor timer method may require more iterations but is expected to eventually
converge on a more balanced set of processor boundaries.
Although on the surface the two appear to be radically different approaches,
they can be combined in such a way as to obtain both the quick convergence of particle balancing and the superior converged solution of processor timing.  We call this hybrid method a timer-augmented cost function (TACF).

\begin{figure}
	\begin{center}
		\includegraphics[width=4.5in]{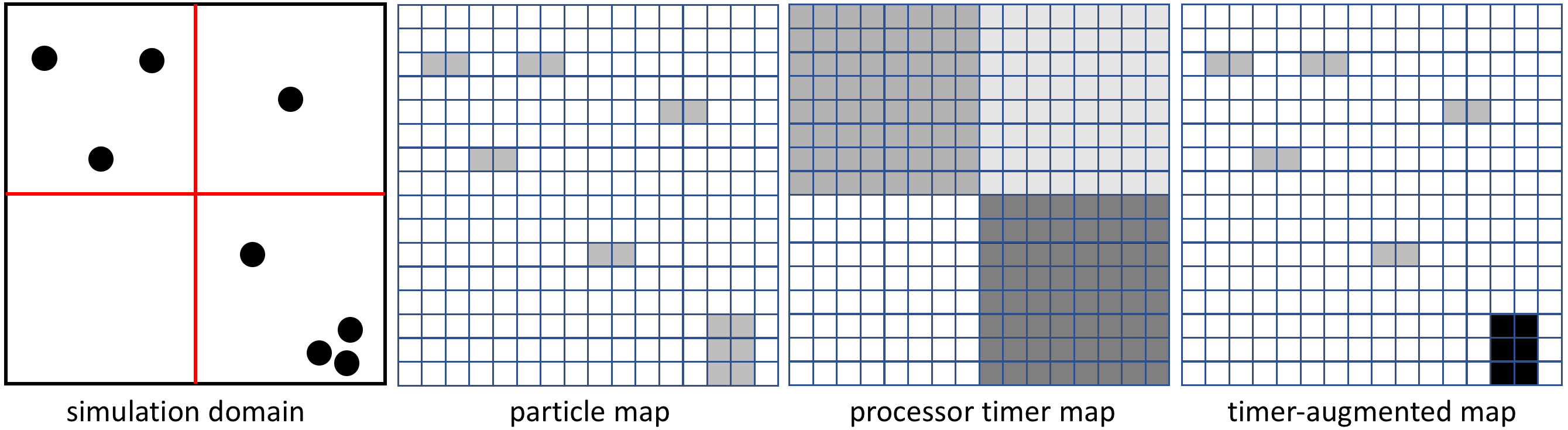}
	\end{center}
	\caption{Schematic of various methods for generating a cost map.  The simulation domain is shown on the left, with particles in black and processor boundaries in red.  The other figures are cost maps, with darker shades indicating higher estimated cost (scale differs between maps).  Particle balancing produces a map which is proportional to particle density.  Processor timers best reflect the true cost of the computation, but are blind to the distribution of load within each processor.  The timer-augmented map resolves the spatial distribution of particles while also reflecting the extra cost associated with, e.g., particularly dense gas.}
	\label{fig-mapschematic}
\end{figure}

To show how this can be done, we first sketch the process of building a cost map (Fig. \ref{fig-mapschematic}) for a particle balancer.  A cost map is just the discretized cost function.  We take a grid spanning the entire simulation domain, where each cell has only a single variable: number of particles.  Each processor will go over all of its particles, determine in which grid cell each belongs, and increment that cell's counter. In the end we have a high-resolution map of particles for the whole domain, which will be given to the partitioner.

However, there is no reason that every particle should contribute the same
  weight (i.e., same estimated computational load) to the map.  If
performing a multi-species simulation, where one class of particle is
significantly more expensive to move or collide than another, a user or developer might choose for
these particles to have a disproportionate effect on the cost map -- perhaps each cell's counter is incremented by 2 instead of 1 for these expensive
particles. Then when the partitioner operates on the cost map, it will give fewer total particles to processors which end up with more of the expensive particles.

Our insight is that we can instead (or in addition) weight particles' contributions to the cost map by processor-level timer results.  If a processor containing $N$ particles took $T$ seconds to compute the previous time step, then when it contributes to the cost map it will not add 1 to each cell per particle, but instead $T/N$.  In total, it will contribute $T$ to the cost map, distributed across the cells that overlap its subdomain in proportion to where its particles are located.  When every processor does this, each with its own values of $T$ and $N$, the sum across all cells in the resulting map is just the total time taken by all processors for the previous time step.  The amount of this quantity in a region of space is an estimate of the time required to compute a time step for that region.  When the partitioner evenly apportions this quantity, each processor will end up with an amount of ``time" which is closer to the average.  We are essentially taking a reasonably good performance model that nevertheless has some systematic error and augmenting it with processor-level timers to drive its converged error to zero.  Where particle balancing converges on an even distribution of particles (and therefore an almost-even distribution of computational load), this timer-augmented method just converges on an even distribution of computational load.

This augmented cost model can also be conceived of starting from the timers.  If all we have are processor-level timers, then we must guess at a distribution of the load within each processor.  We may assume that it is uniform for lack of better options.  Instead, we can suppose that the load is distributed in the same way as the particles, and this will work fairly well because particle count is a reasonably good proxy for load.  This does not change the converged result, since eventually all processors should be taking equal amounts of time either way.  What we gain is faster convergence.  Imagine a processor which takes twice as long as the average processor, and which contains a highly non-uniform particle distribution.  All of the particles are in just one half of its volume.  If our partitioner is blind to the distribution of the load inside each processor, it might re-draw boundaries in useless or counter-productive ways.  It might donate the empty half of the over-worked processor's volume to other processors, achieving nothing.  It might donate the half containing all of the particles, leaving the original processor with nothing to do and over-working another processor.  Such a system can be made to eventually converge, but it will require multiple iterations.  And we stress that these iterations are expensive.  Not only does a balancing pass itself require a great deal of communication, but the simulation cannot simply repeatedly build cost maps and partition processors, since the system needs timer data for the new processor distribution to perform a new balancing pass.

\section{Method}
We implemented a load balancer with the ability to produce a cost map using
particle balancing, processor timers, and with our TACF in a simple 3D DSMC code.  The base code essentially follows along with the treatment in \cite{boyd2017nonequilibrium}.  It is parallelized with MPI.

To determine processor boundaries, we use a cost map consisting of a uniform grid with 1000 cells per processor, with the cells distributed so as to keep the map's resolution similar in all three dimensions.  At initialization we set the cost map to a constant value everywhere, and then we periodically produce new cost maps from the particles and/or timer results of the ongoing simulation.  After producing a cost map, we use the recursive coordinate bisection algorithm to obtain boundaries.  Essentially, this algorithm identifies the longest dimension in the domain, then makes a cut in the plane normal to it so that half of the cost is on one side and half on the other (it interpolates within grid cells).  Then it operates recursively on the two subdomains which were just created by the cut, until the domain has been split into $2^n$ subdomains with estimated equal cost, suitable for $2^n$ processors.  These cuts are stored in a tree, and the tree can then be traversed to find which processor owns a given particle by checking the particle's position against the cut position at each of $n$ levels.  This is a simple technique that only works for processor counts which are powers of two, but we believe our results would also apply to other partitioning methods.

Simulations were run on Intel Xeon E5-2680 v3 Haswell CPUs, with two CPUs (16 cores total) and 128 GiB of memory per node.  The authors gratefully acknowledge the computing time granted through JARA-HPC on the supercomputer JURECA at Forschungszentrum J\"ulich\cite{jureca}.

\subsection{Test Case}\label{sec:test-case}
\begin{figure}
	\begin{center}
		\includegraphics[width=4.5in]{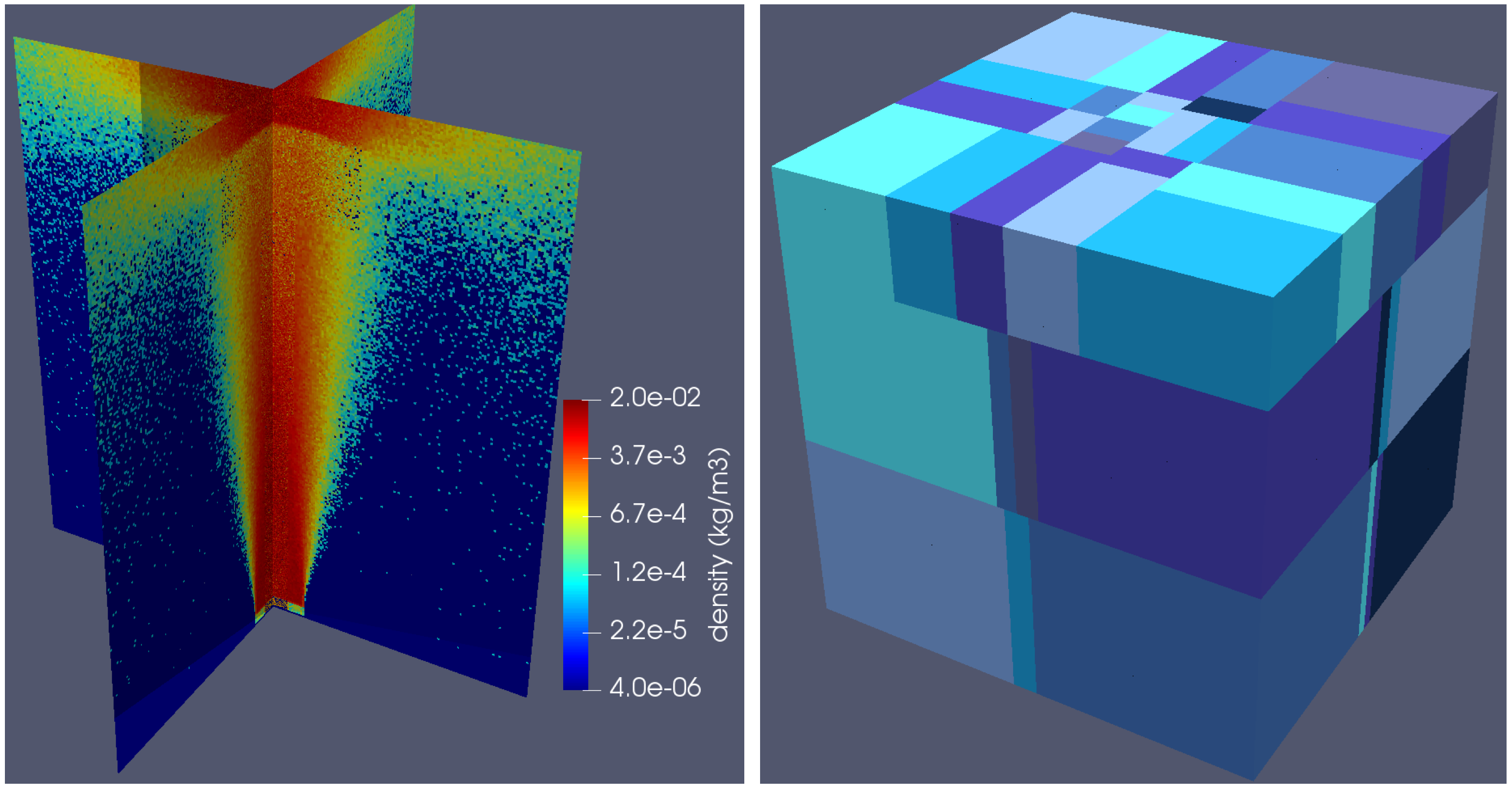}
	\end{center}
	\caption{(left) Contours of density along two planes which cut through the center of the domain.  (right) A sample processor decomposition for our test case at steady state, using particle balancing with 64 processors.}
	\label{fig-domain}
\end{figure}

Our test case is an argon jet shooting upwards into a vacuum towards a solid
wall, shown in Fig. \ref{fig-domain}.  The domain is a cube with side
length 80 cm.  Each time step, argon flowing upwards at 2900 m/s (Mach 9) is created in a cylinder on the bottom boundary at 0.01
kg/m$^3$ and 300 K.

The resulting argon jet starts at the bottom of the domain and expands upwards, becoming faster, colder, and less dense.  A strong shock forms just off of the solid boundary at the top of the domain, which specularly reflects incoming molecules.  The gas behind this shock is much denser and hotter, and is now subsonic.  It then accelerates towards the vacuum boundaries on the sides of the domain.

This case features typical spatial non-uniformity.  Density varies by many
orders of magnitude, temperatures range from nearly 0 to more than 5000 K, and
there is both supersonic and subsonic flow.  Molecules are created each time
step in a small region of the domain, and there are both vacuum and solid
boundaries.  Further, only the boundary conditions and creation mechanism are
known a priori.  The flow-field in Fig. \ref{fig-domain} is the eventual
steady state of a time-varying 3D DSMC simulation.  In short, it would be very hard to specify an appropriate domain decomposition in advance, and the optimal decomposition will change as the flow develops from an initially-empty domain.

Fig. \ref{fig-domain} also shows an example of how the problem is partitioned.  Large processor subdomains are placed over the nearly empty bottom corners of the domain while many small processors are clustered over the domain's centerline, where most of the molecules are.

To make our findings reproducible, we now detail the various parameters we chose for our simulations. 
We use a time step of $1.427 \times 10^{-7}$ s.  The ratio of real molecules
to simulated molecules is $2.4 \times 10^{12}$ per processor, and each
processor has roughly 100,000 cells, distributed in as close to a regular,
uniform grid as possible within its subdomain.  That is, as the number of
  processors increases, the physical flow being simulated is unchanged, but we use more computational molecules and more cells -- this is very close to weak scaling.  Processor boundaries are initialized to be uniform, and for the first 50 time steps argon is created at 1\% of its nominal density (so as not to run out of memory on the small number of processors which own the creation region before any load-balancing occurs).  We load balance every 25 time steps for the first 100 time steps, then every 50 time steps thereafter, or when any processor owns more than 4 million molecules (again to avoid running out of memory).  The simulation reaches steady state around the 600th or 700th time step.  After the 900th time step, we stop load balancing and run for 100 more time steps.  The frequency of load-balancing is not optimized here, and optimizing this is itself a difficult problem, which is why we mainly focus on the balance at steady state.

\section{Results}\label{sec:results}
We can directly compare particle balancing to TACF 
by looking at the quality of the decomposition each produces.  We obtain the
wall clock time for a time step by starting a timer just after molecules are
communicated (after they move and potentially arrive in other processors'
subdomains) and stopping it at the same point in the next time step.  We
obtain a ``processor time" for a single processor (that is, a core) for a time step by starting
a timer just after molecules are communicated and stopping it just before they
are communicated in the next time step.  This measure is capturing nearly all
of the computation that a processor is doing while excluding time that it
spends waiting for other processors.
  This is the quantity that each processor contributes to the cost map and is what we attempt to balance with the aim of minimizing the wall clock time.

\begin{figure}
	\centering
	\begin{tikzpicture}[scale=0.8]
	\begin{axis}[
	ybar stacked, bar width=10pt, bar shift = -6.5pt,
	xmin=-.5, xmax=10.5,
	ymin=0, ymax=.6,
	width=14cm,
	height=5cm,
	xtick={0,1,2,3,4,5,6,7,8,9,10},
	xticklabels={1,2,4,8,16,32,64,128,256,512,1024},
	legend style={at={(0.5,-0.3)},anchor=north,/tikz/every even column/.append style={column sep=0.5cm}, anchor=north,draw=none, legend columns=4},
	ymajorgrids,
	ylabel={Time (s)},
	axis x line*=bottom,
	y axis line style={opacity=0},
	xlabel ={Number of Processors},
	title={Mean and Wall Clock Time}
	]
	\addplot[color=black, fill=blue!90!green] table[x expr=\coordindex, y index=0, header=false]{img/particle.txt};
	\addplot[color=black, fill=blue!30!green] table[x expr=\coordindex, y index=1, header=false]{img/particle.txt};
	\legend{mean processor time, wall clock time}				
	\end{axis}
	\begin{axis}[
	ybar stacked, bar width=10pt, bar shift = 6.5pt,
	xmin=-.5, xmax=10.5,
	ymin=0, ymax=.6,
	width=14cm,
	height=5cm,
	xtick={0,1,2,3,4,5,6,7,8,9,10},
	xticklabels={1,2,4,8,16,32,64,128,256,512,1024},
	y axis line style={opacity=0},				
	]				\pgfplotsinvokeforeach{0,1,2,3,4,5,6,7,8,9,10}{
		\node[anchor=south, white, scale=0.6] (A) at ($ (axis cs:#1,0.0) + 1*(6.5pt, 0)$) {\sc H};
		\node[anchor=south, white, scale=0.6] (A) at ($ (axis cs:#1,0.0) - 1*(6.5pt, 0)$) {\sc P};
	}
	\addplot[color=black, fill=blue!90!green] table[x expr=\coordindex, y index=0, header=false]{img/hybrid.txt};
	\addplot[color=black, fill=blue!30!green] table[x expr=\coordindex, y index=1, header=false]{img/hybrid.txt};
	\end{axis}

	\end{tikzpicture}				
	\vspace{-0.5cm}
	\caption{Mean processor and wall clock times for particle balancing (P) and the hybrid timer-augmented method (H).  The mean processor time is the average of all of the individual processor times used as inputs for the timer-augmented load balancer.  The wall clock time is the actual time required for the simulation to complete a time step.}
	\label{fig-methodcompare}
\end{figure}
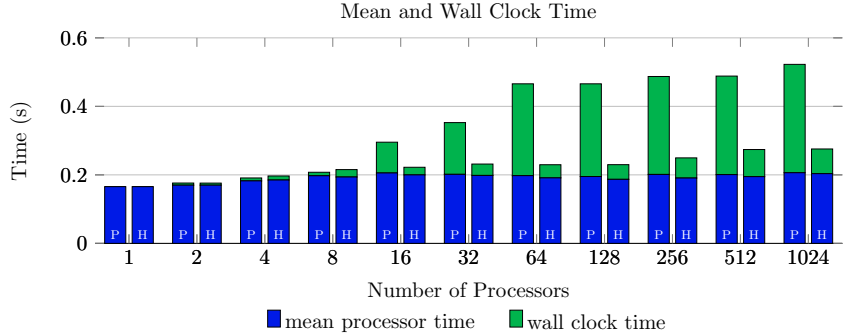

Fig. \ref{fig-methodcompare}  shows the mean processor
times and wall clock times for particle and timer-augmented balancing for a range of
processor counts, with the problem scaled to match per Section \ref{sec:test-case}.  Measurements were taken over the final 50 time steps of the simulation.  The mean processor times are shown to provide a baseline.  Wall clock times in excess of the mean processor times are due to either imbalance or communication overhead.

Up to 8 cores, both methods perform well -- there is little excess wall clock time.  The test case is symmetric on two of the coordinate axes, and the partitioner makes its first cuts parallel to the planes in Fig. \ref{fig-domain}, so even particle balancing produces four mirrored subdomains which all do basically equal amounts of computation.  A particle balancer can get lucky like this, where a non-uniformity in one processor is balanced by a similar non-uniformity in another.  This can even happen without symmetry.  Especially when processor counts are small, it is likely that subdomains are large enough to contain a variety of flow regimes such that expensive regions and cheap regions average out.

However, the particle balancer falls behind the timer-augmented balancer starting at 16 cores.  By 64 cores, the particle balancer is producing a
decomposition where processors are on average spending more time waiting for
other processors than on computation. The inefficiency due to imbalance does not grow without bound, though.  After quickly growing between 16 and 64 cores, it grows only very slowly up to 1024 (and this growth may be largely due to communication costs).  This is predicted by our theoretical discussion of the advantages and disadvantages of different load balancers in Section \ref{sec:state-of-the-art}.  There is a limit to how imbalanced a particle balancer can get, which is determined by how expensive the most expensive particles in the simulation are and how cheap the cheapest particles in the simulation are.

Meanwhile, the TACF balancer performs much better for large processor
counts.  It sees only slow growth in wall clock time as processor count
increases, as might be expected of a weak scaling plot.  The practical benefits are clearly significant -- the simulation using the TACF balancer is able to perform time steps about twice as quickly on many cores.

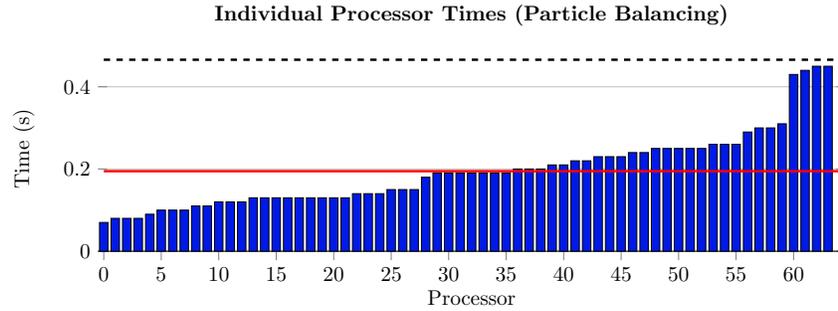
\begin{figure}
	\centering
	\begin{tikzpicture}[scale=0.8]
	\begin{axis}[
	ybar, bar width=4pt,
	xmin=-.5, xmax=64.5,
	ymin=0, ymax=.5,
	width=14cm,
	height=5cm,
	ymajorgrids,
	ylabel={Time (s)},
	axis x line*=bottom,
	y axis line style={opacity=0},
	xlabel ={Processor},
	title={\textbf{Individual Processor Times (Particle Balancing)}}
	]
	\addplot[color=black, fill=blue!90!green] table[x expr=\coordindex, y index=0, header=false]{img/particle_64proctimings.txt};	
	\draw [line width = .4 mm, red] (axis cs: 0,.1947) -- (axis cs:64,.1947);
	\draw [line width = .4 mm, black, dashed] (axis cs: 0,.4658) -- (axis cs:64,.4658);	
	\end{axis}
	
	\end{tikzpicture}				
	\vspace{-0.5cm}
	\caption{Processor times for the particle balancing method.  The red line shows the mean processor time and the dashed line shows the wall clock time..  Deviations from the mean indicate imbalance.}
	\label{fig-64proctimings_particle}
\end{figure}

\begin{figure}
	\centering
	\begin{tikzpicture}[scale=0.8]
	\begin{axis}[
	ybar, bar width=4pt,
	xmin=-.5, xmax=64.5,
	ymin=0, ymax=.5,
	width=14cm,
	height=5cm,
	ymajorgrids,
	ylabel={Time (s)},
	axis x line*=bottom,
	y axis line style={opacity=0},
	xlabel ={Processor},
	title={\textbf{Individual Processor Times (Timer-Augmented Balancing)}}
	]
	\addplot[color=black, fill=blue!90!green] table[x expr=\coordindex, y index=0, header=false]{img/hybrid_64proctimings.txt};
	\draw [line width = .4 mm, red] (axis cs: 0,.1906) -- (axis cs:64,.1906);
	\draw [line width = .4 mm, black, dashed] (axis cs: 0,.2316) -- (axis cs:64,.2316);		
	\end{axis}
	
	\end{tikzpicture}				
	\vspace{-0.5cm}
	\caption{Processor times for the timer-augmented balancing method.  The red line shows the mean processor time and the dashed line shows the wall clock time.  Deviations from the mean indicate imbalance.}
	\label{fig-64proctimings_hybrid}
\end{figure}
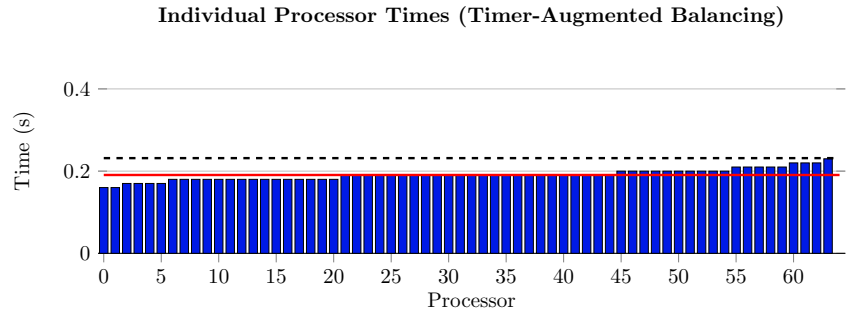

We now look more closely at the efficacy of each load balancer by examining
distributions of processor times.  Fig. \ref{fig-64proctimings_particle} shows
the (sorted) processor times for the 64-core particle-balanced simulation. Most processors are more than
20\% off of the mean time.  The high wall clock time seen earlier is driven by four particularly slow processors which cover the most expensive parts of the flow.  By contrast, the distribution of times for the TACF balancer (Fig. \ref{fig-64proctimings_hybrid}) are much more even, with almost all processors within 10\% of the mean time and no significant outliers.

The TACF method achieves this improved load balance by recognizing that not all particles are equally costly, so it requires more memory.  In all of the particle balancing simulations, there were approximately 650,000 molecules in each processor's subdomain.   While this is true of the average processor with TACF, there is significant particle imbalance starting at 16 processors.  With 16, one processor has 1 million molecules.  On 64, one has 1.44 million.  On 1024, one has 1.8 million.  If memory use is a constraint, TACF could be modified to cap the maximum particle imbalance by finding a minimum alternative weight that particles will contribute to the cost map even if their processor is very fast and contains many particles.

Comparing TACF to the processor timer method is more difficult.  Both should converge to similar decompositions after enough balance passes, but the hybrid method should converge faster.  However, real implementations of processor timer methods make use of sophisticated partitioning schemes with implicit cost functions to try to address this issue, and so a fair comparison is impossible without implementing something similar.  We note that it is hard to do better than TACF at steady state, per Figs. \ref{fig-methodcompare} and \ref{fig-64proctimings_hybrid} -- only a small improvement from better load-balancing is possible.  To try to study the transient performance of each, we can run our test case with processor timer balancing by using a damping factor, such that processors contribute a weighted average of their individual processor times and the mean processor time to the cost map.  When we do this (including tuning the damping factor), the simulation takes significantly longer (at least 2x longer when using more than 16 processors) to complete than with particle balancing or TACF.  Further, we must perform an ten extra load balance passes at steady state to obtain a reasonably well-balanced decomposition with a wall clock time per time step comparable to TACF.  The processor timer method performs poorly during the transient phase of the simulation since it does not make sense to perform many load balancing passes every time a new partition is desired, whereas particle balancing and TACF perform about as well during the transient phase as at steady state.

\section{Conclusion}

Large-scale DSMC simulations require load balancing in order to be feasible.  As part of this, many methods use a model of the load as a function of space to guide the decomposition of the domain into processors' subdomains.  We discussed several models, and proposed a timer-augmented cost function which combines the quick convergence of a particle balancer and the low error of processor timers.

Not only does a timer-augmented cost function yield a significantly more balanced domain decomposition than the one achieved by partitioning on the basis of particles alone, it is also an easy improvement to make in code.  In the case of our code, the only difference between a particle balancer and the TACF balancer is that, with TACF, each processor contributes a different constant value to the cost map instead of the same constant value.  All other aspects of the load balancer and partitioner can remain the same, which makes it easy to realize significant performance gains ($\sim$2x at steady state for our test case).  In fact, a timer-augmented balancing method was recently adopted by the developers of SPARTA, and is now an option for users of this open-source code.

%
%
%
%


\bibliographystyle{splncs04}
\bibliography{mcdoniel_vecpar18}
\end{document}